\documentstyle[aps,preprint]{revtex}
\draft
\begin{document}
\title{Resistivity due to a Domain Wall in Ferromagnetic Metal} 
\author{Gen Tatara}
\address{Graduate School of Science, Osaka University, Toyonaka, Osaka 560, 
Japan}
\author{Hidetoshi Fukuyama} 
\address{Department of Physics, University of Tokyo, Hongo, Tokyo 113, 
Japan}
\date{\today}
\maketitle

\begin{abstract}
The resistivity due to a 
domain wall in ferromagnetic metallic wire is calculated based on the linear 
response theory. 
The interaction between conduction electrons and the wall is expressed in 
terms of a classical gauge field 
which is introduced by the local gauge transformation in the electron spin 
space.
It is shown that the wall contributes to the decoherence of electrons and that 
this 
quantum correction can dominate over the Boltzmann resisitivity, leading to a 
{\it decrease} of resisitivity by nucleation of a wall.
The conductance fluctuation due to the motion of the wall is also investigated.
The results are compared with recent experiments.
\end{abstract}

The interplay between the electron transport properties and 
the magnetic object such as a magnetization and a domain wall(DW) has 
recently been attracting much attention. 
Of particular interest is the case of mesoscopic system where the motion of a 
DW or a magnetization can be driven by quantum fluctuation and 
described as a macroscopic quantum phenomena\cite{Chichilianne}.
In the case of the quantum depinning of a 
DW\cite{Stamp}, for example, a theoretical study\cite{TF} indicates 
that the depinning can be affected by the dissipation 
caused by the conduction electron if the thickness of the wall, $\lambda$, is 
small, e.g., a few lattice constants. 
The change of the magnetization associated with such a depinning of a 
mesoscopic DW is very small and so 
it is very difficult to observe such small magnetic objects directly, e. g.
by SQUID.
The transport properties on the other hand can detect 
a very small change of the magnetization as a change of resistance.
Indeed recently in a mesoscopic wire of Ni with diameter of 300\AA\ 
several small discontinuous changes of the resistivity 
have been observed as the magnetic field is swept\cite{Hong}.
It is argued there that these jumps are due to the change of the total 
magnetization by depinning of a DW, and the displacement of the wall has been
 estimated from the value of magnetoresistance to be 
$\sim 1.2\mu$m\cite{Hong}.
Other possible origins of this jump are proposed in this paper.

Not only in these mesoscopic system
 the interplay between the magnetic structure and the electronic 
transport properties may play important roles in the bulk system; e. g., 
in double exchange systems like La$_{1-x}$Sr$_{x}$MnO$_{3}$, 
scattering by DWs is considered as a possible origin of 
low temperature magnetoresistance\cite{Schiffer}.

In this paper we study the resistivity in ferromagnetic metals arising from the 
scattering by a DW on the basis of linear 
response theory by taking account of the impurity scattering at the same time.
The case of $\lambda \ll l$ ($l$ being the elastic mean free path) has been 
studied 
by Cabrera and Falicov\cite{Cabrera} in the classical Boltzmann approximation.
Their result indicates that the resisitivity becomes large 
only when the spin splitting is comparable to the Fermi energy and 
$k_{F}\lambda\lesssim 1$ ($k_{F}$ being the Fermi momentum).
In thier study, however, electronic motions have been assumed to be 
one-dimensional, which is not realistic, at least at present, in actual 
metallic wires.
Here we study the effect of a DW on resistivity in a mesoscopic wire 
with width $L_{\perp}$ satisfying 
$\lambda \gtrsim L_{\perp} \gg k_{F}^{-1}$, 
thus treating the electron as three-dimensional.
The length is $L$ and the wire 
direction has been  chosen as $z$-axis. 
We investigate the quantum corrections to the resistivity 
by a wall as well as the Boltzmann resistivity. 
The conductance fluctuations\cite{LSF,Feng} arising from the motion of the 
wall has also been calculated. 

We consider explicitly the case described by a single-band Hubbard model 
in the Hartree-Fock approximation\cite{sd}. 
The calculation is carried out at zero temperature. 
The Lagrangian of the electron (denoted by $c^{(0)}$) in the imaginary 
time is given as
\begin{equation}
L=\sum_{{\bf k}{\sigma}}
c^{(0)\dagger}_{{\bf k}{\sigma}}
(\partial_{\tau}+\epsilon_{{\bf k}})
c^{(0)}_{{\bf k}{\sigma}}
-U\sum_{{\bf x}}\mbox{\boldmath{$M$}}(c^{(0)\dagger}
\mbox{\boldmath{$\sigma$}}c^{(0)}), \label{L}
\end{equation}
where 
$\epsilon_{{\bf k}}\equiv 
\hbar^{2}{\bf k}^{2}/2m-\epsilon_{F}$
($\epsilon_{F}$ being the Fermi energy) and $U$ is the Coulomb interaction. 
The spin index is denoted by 
$\sigma=\pm$ and $\mbox{\boldmath{$\sigma$}}$ is the Pauli matrix.
The magnetization vector, $\mbox{\boldmath{$M$}}$, is in unit of Bohr 
magneton.
The configuration of $\mbox{\boldmath{$M$}}$ is determined by the 
ferromagnetic Heisenberg model\cite{TF},
\begin{equation}
H_{M}=\sum_{{\bf x}}\left[ 
\frac{J}{2}|\nabla \mbox{\boldmath{$M$}}|^{2}
-\frac{K}{2}M_{z}^{2} \right],
\end{equation}
where $J$ is the effective exchange energy determined by $U$ 
and $K$ is the magnetic anisotropy energy introduced 
phenomenologically\cite{demag}.
Here we are interested in the solution of a single domain 
wall. In terms of the polar coordinate that represents the direction of 
magnetization vector, $(\theta,\phi)$, the solution of a DW is given by
$\cos\theta=\tanh\frac{z}{\lambda}$ and 
constant $\phi$, where $\lambda=\sqrt{K/J}$.

In Eq. (\ref{L}) the last term represents the interaction between the 
magnetization and the electron. 
For the perturbative calculation of 
resistivity, we need to rewrite this term 
by use of the local gauge transformation in the spin space,
\begin{equation}
c_{\sigma}\equiv 
\sigma\left(\cos\frac{\theta}{2}c^{(0)}_{\sigma}-i\sin\frac{\theta}{2}
c^{(0)}_{-\sigma}\right).
\end{equation} 
In terms of the new electron operator, $c$, the Lagrangian is written 
as\cite{TF}
$L=\sum_{{\bf k}{\sigma}}
c^{\dagger}_{{\bf k}{\sigma}}
(\partial_{\tau}+\epsilon_{{\bf k}\sigma})
c_{{\bf k}{\sigma}}$$
+H_{\rm int}$,
where 
$\epsilon_{{\bf k}\sigma}
\equiv\epsilon_{{\bf k}}-\sigma \Delta$ 
with $\Delta\equiv U |\mbox{\boldmath{$M$}}|$ being half the splitting 
between the up and down spin electrons. The interaction is obtained as 
\begin{equation}
H_{\rm int}=\frac{\hbar^{2}}{2m}\sum_{{\bf k}}\sum_{q//z}
\left( -\left(k_{z}+\frac{q}{2}\right)a_{q}c^{\dagger}_{{\bf k}+q}
\sigma_{x}c_{{\bf k}}
+\frac{1}{4}\sum_{p//z}
a_{p}a_{-p+q}c^{\dagger}_{{\bf k}+q} c_{{\bf k}} \right).\label{Hint}
\end{equation}
Here $a_{q}\equiv  (1/V)\sum_{{\bf x}}e^{-iqz}\nabla_{z}\theta $
$=(\pi/L)e^{-iq z_{i}}[1/{\rm ch}(\pi q\lambda/2)]$
($V\equiv L_{\perp}^{2}L$ and $z_{i}$ being the center 
coodinate of the domain wall).
Due to this gauge transformation, the electronic current in $z$-direction is 
changed to be
$J_{z}=J_{z}^{0}+\delta J$, where 
$J_{z}^{0}\equiv 
({e\hbar}/{m})\sum_{{\bf k}} k_{z}c^{\dagger}_{{\bf k}} c_{{\bf k}}$ and 
\begin{equation}
\delta J\equiv  -\frac{e\hbar}{2m} \sum_{{\bf k},q//z}
a_{q}c^{\dagger}_{{\bf k}+q}\sigma_{x}c_{{\bf k}} .  
\end{equation}

By use of the Kubo formula, 
the conductivity for the current along the wire is calculated from the 
current-current correlation function.
By assuming that the scattering due to normal impurities are dominant, we 
estimate the effect of DW on the correction to the conductivity 
perturbatively to the second order 
of $a$. 
(The first order contribution vanishes.)
The second order contributions to the Boltzmann conductivity
 are shown in Fig. \ref{FIGdiagram}.
The process $Q_{1}$ arises from the correlation of $\delta J$ and $Q_{3}$ 
is due to $\delta J$ and an interaction with the wall.  
$Q_{2}$ and $Q_{4}$ are the self-energy corrections due to the wall 
and $Q_{5}$ is the 
vertex correction to the correlation of $J_{z}^{0}$.
After straightforward calculation and by use of the particle-hole symmetry, 
which we assume, 
$\Delta Q\equiv\sum_{i=1}^{5}Q_{i}$ is shown to be
\begin{equation}
\Delta Q(i\omega_{\ell})=
\frac{1}{2}\left(\frac{e\hbar \Delta}{m}\right)^{2}
\frac{1}{\beta}\sum_{n}\frac{1}{V}\sum_{{\bf k} q\sigma }|a_{q}|^{2}
G_{{\bf k}-\frac{q}{2},n,\sigma}
G_{{\bf k}-\frac{q}{2},n+\ell,\sigma}
G_{{\bf k}+\frac{q}{2},n,-\sigma}
G_{{\bf k}+\frac{q}{2},n+\ell,-\sigma}
.
\end{equation}
Here $\omega_{\ell}\equiv 2\pi\ell/\beta$ and the Green function is given by
$G_{{\bf k},n,\sigma}\equiv
 1/[i(\varepsilon_{n}+
 (\hbar/2\tau){\rm sgn}(n))-\epsilon_{{\bf k}{\bf k}{\bf k}{\bf k}\sigma}]$,
where $\varepsilon_{n}=\pi(2n+1)/\beta$ and $\tau$ is the life time due to 
the normal impurity scattering and ${\rm sgn}(n)=1$ and $-1$ for $n>0$ 
and $n<0$, respectively.

Hence the correction to the Boltzmann conductivity by a DW,
$\Delta \sigma$, is obtained as $\Delta \sigma=-\sigma_{0}A$
where
$\sigma_{0}\equiv e^{2}n\tau/m$ ($n$ being the electron density) 
is the Boltzmann conductivity without the wall and $A$ is given by 
\begin{equation}
A=\frac{\pi}{\hbar}\frac{\Delta^{2}\tau}{2nV}n_{\rm 
w}\sum_{\sigma,\pm}\frac{N_{\sigma}}{k_{{\rm F}\sigma}}
\int_{0}^{\infty}\frac{dx}{x}
\frac{1}{{\rm ch}^{2}x}
\tan^{-1}\left(\frac{2l_{\sigma}}{\pi \lambda}x\pm 2\Delta
\frac{\tau}{\hbar}\right).
\label{Aexpression}
\end{equation}
Here $l_{\sigma}\equiv \hbar k_{F\sigma}\tau/m$,
$n_{\rm w}\equiv 1/L$ being the density of the wall
and $N_{\sigma}\equiv (mk_{F\sigma}V/2\pi^{2}\hbar^{2})$ is the 
density of states at the Fermi energy of the electron with spin $\sigma$.
The wall contribution to the resisitivity is given as $\rho_{\rm 
w}\equiv\sigma_{0}^{-1}((1-A)^{-1}-1)\simeq \sigma_{0}^{-1}A$.

We consider a ferromagnet where $\Delta \tau /\hbar\gg 1$ is satisfied. 
Then Eq. (\ref{Aexpression}) reduces to 
\begin{equation}
A \simeq \frac{3n_{\rm w}}{2m n\lambda }\sum_{\sigma}
\frac{N_{\sigma}}{V}
.\label{Avalue}
\end{equation}

Let us look into the effect of the wall on qunatum transport properties in 
disordered system, where the interference effect, which is 
represented by the maximally-crossed diagram (Cooperon), becomes important.
The processes which describe the effect of the wall on the qunatum 
correction at low energy are shown in Fig. \ref{FIGladder}.
They both contribute to the dephasing of the electron, but 
the vertex type process (b) 
includes Cooperons which connect the electrons with differrent spin, 
and thus is suppressed  in ferromagnets we are considering 
due to the condition $\Delta\tau/\hbar\gg 1$. 
Hence only the 
self-energy type (a) is important here.
The higher order contributions similar to this process can be 
summed up  giving rise to the mass of the Cooperon. 
The quantum correction by the wall is then obtained as 
\begin{equation}
\sigma_{Q}=\frac{2\hbar e^{2}{k_{F}^{2}}\tau}{3\pi m^{2}} 
\frac{1}{V}\sum_{q}
\left(\frac{1}{Dq^{2}}-\frac{1}{Dq^{2}+\frac{1}{\tau_{\rm w}}}\right),
\end{equation}
where $D\equiv k_{F}^{2}\tau/3$ and 
$\tau_{\rm w}$ is the life-time due to the wall given by 
${1}/{\tau_{\rm w}}
          =(n_{\rm w}{k_{F}^{2}}\hbar^{4}/{6\Delta^{2}\lambda\tau})$.
In the case where $DL_{\perp}^{-2}>\tau_{\rm w}^{-1}$, which we assume, 
the $q$-summation should be carried out along the one-dimention
with a cut off of $L^{-1}$ for small $q$. 
The result for $L/l \gg 1$ and $\kappa\equiv \tau/\tau_{\rm w}\ll 1$ is
\begin{equation}
\frac{\sigma_{Q}}{\sigma_{0}}\simeq \frac{6}{k_{F}^{2}L_{\perp}^{2}}
\left(\frac{L}{l}-\frac{\tan^{-1}(\sqrt{3\kappa}L/l)}{\sqrt{3\kappa}}
\right).
\label{sigmaQ}
\end{equation}
Note that $\sigma_{Q}$ is positive, sunce the DW 
suppresses the interference due to random impurity scattering.

So far we have studied a static wall.
Let us now discuss the conductance fluctuation due to the motion of the wall.
In this case a small jump of a wall can result in substantial 
change in resistivity, in contrast to the change due to the effect of 
classical magnetoresisitance\cite{Hong}, 
which becomes important only when the wall moves 
over a distance comparabel to $L$. 
The calculation goes in the siminlar way as the conductance change due to the 
motion of a single atom in a disordered metal\cite{Feng}.
The square of the conductance change $\delta G$ 
due to the motion of a wall over a distance of $r$ is 
evaluated by calculating the diagram with two bubbles 
with the wall position at $z=r$ and $z=0$ connected by impurities and the wall. 
A typical diagram is shown in Fig. \ref{FIGucf}. 
The DW line here represents the motion of the wall and Cooperons 
include the mass arising from the wall, $1/\tau_{\rm w}$. 
There are other diagrams with the contribution of the same order 
which contains one or two more impurity 
ladders\cite{LSF} and the result of $\delta G$ is obtained as
\begin{equation}
\frac{\delta G(r)}{e^{2}/h}
=\sqrt{2}\frac{4\pi}{3}  \epsilon^{2}
\kappa \alpha
 \times\left\{
   \begin{array}{cc}
      \frac{1}{6}\left(\frac{r}{\lambda}\right)^{2}
 & (r\ll l, \lambda) \\
     1  &  (r\gg l, \lambda)
    \end{array}
    \right. ,
    \label{deltaG}
\end{equation}
where $\epsilon\equiv l/L$ and 
$\alpha\equiv [\sum_{q}(Dq^{2}\tau+\kappa)^{-4}]^{(1/2)}$ is calculated for 
$\epsilon,\kappa \ll 1$ as
\begin{equation}
\alpha
\simeq
\frac{9}{2\pi}\frac{1}{\epsilon \kappa} \left( 
\frac{5}{24\kappa^{2}}
  \left(\frac{\tan^{-1}(\epsilon/\sqrt{3\kappa})}{\sqrt{3\kappa}}
  -\frac{\epsilon(\epsilon^{2}+5\kappa)}{(\epsilon^{2}+3\kappa)^{2}}\right) 
  -\frac{\epsilon}{(\epsilon^{2}+3\kappa)^{3}}   \right) .
\end{equation}

Let us give an numerical estimate of our theoretical conclusions. 
Consider a wire of Ni or Fe with $L=10\mu$m and $L_{\perp}=300$\AA, where  
$\lambda\sim 500$\AA\cite{Hong}. 
If we consider  $d$ electron ($k_{F}^{-1}\sim 1.5$\AA, 
$\Delta/\epsilon_{F}\sim0.2$) and 
choose $l\sim 1000$\AA, then $\Delta \tau=150$ and Eq. (\ref{Avalue}) leads 
to a very small Boltzmann contribution of 
$A\simeq 1.4\times 10^{-8}$.
For $s$ electron, $\Delta/\epsilon_{F}$ will be smaller by a factor of about 
$10^{-2}$ and then from Eq. (\ref{Aexpression}) we obtain 
$A\simeq 2.7\times 10^{-9}$.
On the other hand, the quantum correction becomes larger for the above values 
of parameters ($\kappa$ is $\sim 3.7\times 10^{-4}$) ; 
$\sigma_{Q}/\sigma_{0}= 1.6\times 10^{-3}$.
Thus DWs will contribute to a decrease of resisitivity 
in a ferromagnetic wire of transition metals. 
If the wall moves over a distance of $r\sim 100$\AA\ in this situation, the 
expected 
conductance change is $\delta G\simeq 5.0\times10^{-3}(e^{2}/h)$. 

In the experiment on Ni\cite{Hong}, a descrete increase of resisitivity of about 
0.2\% 
($\delta\rho\simeq 2\times 10^{-9}\Omega$cm 
or $\delta G\simeq 5\times 10^{-3}(e^{2}/h)$) has been observed 
as the magnetic field is swept above the coercive field, at which
the minimum of resistivity appears. 
Comparison with our study may suggest two possibilities for the cause. One is 
that $\delta\rho$ might be due to the annihilation of a wall. 
The other is that $\delta\rho$ can be the fluctuation due to a motion of wall 
over a distance of $r\sim 100$\AA.
Further studies are needed to determine which is the true origin.
In this context it is interesting to note that a recent experiment on Fe wire 
with width of 3000\AA\  has disclosed the existence of a negative jump of 
$\rho$ followed by a positive one close to the field where $\rho$ becomes 
minimum\cite{Otani}.
This result may suggest that the jumps are due to the nucleation and 
subsequent annihilation of a wall.

Our study indicates that not only in mesoscopic case 
the magnetoresisitance of bulk ferromagnets can be affected by the 
dephasing effect by DWs in particular close to the coercive field, 
where $M=0$ and thus the system will contain many domains.

To summarize, the resistivity arising from the scattering of the conduction 
electron by a domain wall in a wire of ferromagnetic metal is calculated based 
on the linear response theory.  The interaction with the wall is 
expressed as a classical gauge field acting on the electron, which we 
examined in the second order perturbation theory. 
In addition to the Boltzmann resisitivity, we have investigated the effect of 
the wall on quantum transport properties in disordered metals. 
The wall suppresses the intereference between the electron, and hence decreases 
the resisitivity in the weakly localized regime. 
It will be interesting to observe in magnetic wires this reduction of resistivity 
by the nucleation of domains in more definitive ways.
It has been shown that small motion of a wall can lead to substantial 
conductance fluctuation.
The present calculation provides a first quantitative estimate of 
the effect of a domain wall on the mesoscopic transport properties,  
which, we hope, will be useful in the
interpretation of the experimental results in the near future.

Authors thank K. Kuboki, H. Kohno, Y. Otani and K. Takanashi 
for valuable discussions. G. T.  thanks 
The Murata Science Foundation for finantial support.

\begin{figure}
\caption{The contributions to the Boltzmann conductivity 
which are the second order with 
respect to the interaction with the domain wall, denoted by wavy lines. Solid 
lines indicate the electron Green functions and the current vertex 
(expressed by crosses)
with wavy line represents $\delta J$.
\label{FIGdiagram}}

\caption{(a): The dominant process to the qunatum correction of the 
conductivity.  
Hatched square denotes the particle-particle ladder (Cooperon) 
due to the impurity scattering.   Process (b), which contains Cooperons
connecting the electrons with different spin (denoted by $\sigma$ 
and $-\sigma$), 
is unimportant in ferromagnets due to the present condition
$\Delta\tau/\hbar\gg1$.
 \label{FIGladder}}
 
\caption{ An example of diagrams which contributes to the conductance 
fluctuation due to the motion of the wall. Wavy lines represent the motion of
 the domain wall and the Cooperons here include the mass due to the 
domain wall.
\label{FIGucf}}

\end{figure}

\begin{thebibliography}{99}
\bibitem{sd}
In the case with multi-bands like $s$-$d$ model, the results turns out to be  
valid if $U$ in the Hubbard model is replaced by the $s$-$d$ coupling.
\bibitem{demag}
The choice of $z$-axis as an easy axis is due to the demagnetization field
in the case of the wire geometry. The transverse anisotropy energy is 
neglected 
here, since it does not affect our calculation of a static wall.
\end{thebibliography}
\end{document}